\def\papertitle{Sound Effects Dataset Unification With the Universal Category System}
\def\paperauthorA{Jun Woo Beck}
\def\paperauthorB{Alexander Lerch}

\documentclass[twoside,a4paper]{article}
\usepackage{etoolbox}

\usepackage[print]{dafx26v3}

\usepackage{amsmath,amssymb,amsfonts,amsthm}
\usepackage{siunitx}
\usepackage{euscript}
\usepackage[T1]{fontenc}
\usepackage[utf8]{inputenc}
\usepackage{ifpdf}
\usepackage[english]{babel}
\usepackage{caption}
\usepackage{subfig} 
\usepackage{color}
\usepackage{booktabs}
\usepackage{placeins}  

\usepackage{microtype}
\usepackage{paralist}
\input glyphtounicode
\ninept

\newcounter{numauth}
\newcounter{listcnt}
\newcommand\authcnt[1]{\ifdefined#1 \stepcounter{numauth} \fi}

\newcommand\addauth[1]{
\ifdefined#1 
\stepcounter{listcnt}
\ifnum \value{listcnt}<\value{numauth}
\appto\authorslist{, #1}
\else
\appto\authorslist{~and~#1}
\fi
\fi}
\authcnt{\paperauthorB}
\authcnt{\paperauthorC}
\authcnt{\paperauthorD}
\authcnt{\paperauthorE}
\authcnt{\paperauthorF}
\authcnt{\paperauthorG}
\authcnt{\paperauthorH}
\authcnt{\paperauthorI}
\authcnt{\paperauthorJ}
\def\authorslist{\paperauthorA}
\addauth{\paperauthorB}
\addauth{\paperauthorC}
\addauth{\paperauthorD}
\addauth{\paperauthorE}
\addauth{\paperauthorF}
\addauth{\paperauthorG}
\addauth{\paperauthorH}
\addauth{\paperauthorI}
\addauth{\paperauthorJ}

\usepackage{times}
\newif\ifpdf
\ifx\pdfoutput\relax
\else
   \ifcase\pdfoutput
      \pdffalse
   \else
      \pdftrue
   \fi
\fi

\ifpdf 
  \usepackage[pdftex,
    pdftitle={\papertitle},
    pdfauthor={\authorslist},
    pdfsubject={Proceedings of the 29th International Conference on Digital Audio Effects (DAFx26)},
    colorlinks=false, 
    bookmarksnumbered, 
    pdfstartview=XYZ 
  ]{hyperref}
  \usepackage[pdftex]{graphicx}
\else 
  \usepackage[dvips]{epsfig,graphicx}
  \usepackage[dvips,
    pdftitle={\papertitle},
    pdfauthor={\authorslist},
    pdfsubject={Proceedings of the 29th International Conference on Digital Audio Effects (DAFx26)},
    colorlinks=false, 
    bookmarksnumbered, 
    pdfstartview=XYZ 
  ]{hyperref}
\fi
\usepackage[hypcap=true]{caption}
\title{\papertitle}

 \affiliation
 {\paperauthorA\ and \paperauthorB}
 {\href{http://www.musicinformatics.org/}{Music Informatics Group} \\ Georgia Institute of Technology \\ Atlanta, GA,  USA\\
 {\tt \href{mailto:jbeck85@gatech.edu}{jbeck85@gatech.edu}}
 {\tt \href{mailto:alexander.lerch@gatech.edu}{alexander.lerch@gatech.edu}}
 }

\hypersetup{hidelinks, pdfborder={0 0 0}, colorlinks=false}
\begin{document}
\ifpdf 
  \DeclareGraphicsExtensions{.png,.jpg,.pdf}
\else  
  \DeclareGraphicsExtensions{.eps}
\fi


\maketitle

\begin{abstract}
Sound effects (SFX) datasets and libraries often employ distinct tagging schemes, taxonomies, and metadata structures.
This creates challenges for research on SFX classification and generation because incompatible taxonomies lead to siloed datasets that might require individualized approaches, result in non-comparable outcomes, and prevent data merging strategies.
We propose a modular dataset relabeling framework that adopts the Universal Category System (UCS), an industry-standard hierarchical taxonomy for sound effects, as a shared structural foundation.
This open-source framework enables us 
\begin{inparaenum}[(i)]
	\item to convert tags of existing datasets to UCS with a rule-based multi-stage pipeline and conflict resolution to achieve high automatic conversion rates,
	\item to suggest a stratified dataset split for the new labels, and 
	\item to combine multiple datasets.
\end{inparaenum}
To showcase the practical utility, we introduce the \textit{EnvSound-UCS} dataset, a publicly available unified UCS-compliant dataset of environmental sounds with $58{,}057$ sound clips from three sources: AudioSet, FSD50K, and ESC-50.
\end{abstract}

\section{Introduction}
\label{sec:intro}
The rapid growth and ever-increasing complexity of modern audio-focused artificial intelligence and machine learning approaches results in an increased need for large-scale audio datasets.
While there is a variety of datasets available for machine learning tasks in the audio domain, each dataset tends to employ its own tagging vocabulary, metadata schema, and data partitioning strategy.
This creates three practical challenges when working with multiple datasets:
\begin{compactenum}[(i)]
	\item \textit{compatibility \& interoperability}: as the datasets use different, possibly inconsistent taxonomies and vocabularies, neither a meaningful comparison nor the combination of datasets into a larger dataset is possible, and different folder structures, file formats, and ways to store annotations complicate handling a large number of datasets,
	\item \textit{extensibility}: inconsistent, and in some cases ill-documented annotation methodologies make the extension of existing datasets tedious and complicated to achieve,
	\item \textit{coherence}: without a shared category space, extracting semantically consistent subsets from large datasets or assessing coverage gaps across datasets is not straightforward.
\end{compactenum}

There exist efforts to address these dataset challenges for a variety of tasks from the data side, from the data loader side, or from the model side.
Liu et al.\ combined various datasets for piano transcription into the larger dataset ACPAS~\cite{liu_acpas_2021}.
Bittner et al.\ proposed a solution at the data loading stage with a software capable of loading annotations and data from a variety of datasets into a consistent format~\cite{bittner_mirdata_2019}.
Ma and Lerch attempted to address the diversity of annotations through training a generalized representation for the classification of effects that can be relatively easily adapted to a multitude of annotation schemata~\cite{ma_representation_2022}.
The challenges of transparency and documentation have inspired proposals to utilize datasheets, a standardized documentation framework covering motivation, composition, acquisition, pre-processing, intended uses, distribution, and maintenance~\cite{gebru_datasheets_2021}.

Sound effects (SFX)---audio content other than speech and music that is used in media production to support storytelling, convey environments, and enhance narrative immersion~\cite{moffat_unsupervised_2017, sonnenschein_sound_2001}---encompass an exceptionally broad acoustic domain, from environmental ambiences and Foley recordings to designed synthetic textures.
Unlike speech or music, whose classification systems have been standardized for over a century (e.g., the International Phonetic Alphabet~\cite{international_phonetic_association_handbook_2021}), no comparable consensus existed for SFX until the recent adoption of the Universal Category System~\cite{noauthor_universal_2026}.
Professional sound libraries have consequently relied on proprietary naming schemes, while academic datasets have independently developed their own ontologies.
{Cano et al.~\cite{cano_sfx_2004} proposed an MPEG-7 and WordNet-based scheme for SFX library management, observing that production-scale taxonomies require thousands of categories that do not follow a clean hierarchical structure.
This fragmentation between production workflows and research communities makes SFX particularly noteworthy in the context of the dataset challenges outlined above.

The incompatibility of existing SFX dataset annotations is evident across prominent datasets.
Datasets such as FSD50K~\cite{fonseca_fsd50k_2022} (51{,}197~clips, 200~labels), AudioSet~\cite{gemmeke_audio_2017} (527~classes), ESC-50~\cite{piczak_esc_2015} (50~classes, 2{,}000~clips), UrbanSound8K~\cite{salamon_dataset_2014} (10~classes, 8{,}732 clips), and Clotho~\cite{drossos_clotho_2020} (caption-based) each employ distinct labeling conventions: a dog barking sound is labeled \texttt{dog\_bark} in ESC-50, \texttt{Bark} in FSD50K, and \texttt{Dog} in AudioSet---three different representations of the same acoustic event.
Stamatiadis et al.\ proposed SALT~\cite{stamatiadis_salt_2024}, a standardized label taxonomy for unifying audio event datasets with a flat label set.
{Anastasopoulou et al.\ introduced the Broad Sound Taxonomy (BST)~\cite{anastasopoulou_bst_2024}, a general-purpose two-level hierarchy with 5~top-level classes---including music, speech, and sound effects---and 23~second-level classes. As BST treats sound effects as a single top-level class, it lacks the granularity needed for SFX-specific annotation.

It is noteworthy, however, that the diversity and inconsistency of annotations in commercial libraries has driven the industry, independently of academic efforts, to converge on the Universal Category System (UCS)~\cite{noauthor_universal_2026} as the dominant unified standard in sound design and post-production.
UCS defines a hierarchical structure comprising 82~categories, 453~subcategories, and $9{,}972$~synonyms, and has been widely adopted by major commercial sound libraries, enabling direct compatibility between research and production workflows.
 
Despite this widespread adoption in professional workflows, UCS has not found much traction in academic settings.

We propose to leverage this de-facto annotation standard and utilize publicly available data to introduce a pipeline to convert existing, proprietarily labeled SFX datasets to UCS compliant, compatible, and interoperable datasets.
By establishing a shared label space, the framework enables researchers to combine multiple SFX sources into larger training corpora, extract focused subsets for targeted experiments, and seamlessly incorporate new datasets---effectively treating SFX datasets as interoperable modules rather than fixed, monolithic units.
The main contributions of this work are:
\begin{compactenum}[(i)]
	\item a rule-based tag-to-UCS conversion framework with high automatic conversion rates minimizing the need for human interaction,
	\item a UCS-aware data merge and split tool for ensuring stratified splits and multi-source merging,
	\item a new unified UCS compliant dataset \textit{EnvSound-UCS} integrating three existing datasets.
\end{compactenum}

The source code, the converted datasets, and the new combined dataset are publicly available.\textsuperscript{\ref{fn:repos}}

\section{Proposed framework}
\label{sec:framework}

The goal of this work is to design a reliable conversion pipeline for existing SFX datasets with reduced human intervention by utilizing the synonyms defined by UCS. The main objectives are
\begin{inparaenum}[(i)]
	\item to seamlessly integrate multiple existing datasets with incompatible annotations by unifying the label taxonomy,
	\item to allow for easy extensibility of the new dataset, and
	\item to take advantage of the hierarchical taxonomy to allow for easy subsetting of the combined data. 
\end{inparaenum}

For instance, this framework enables compiling a dataset for an animal sound classifier by combining ESC-50's animal classes, FSD50K's \texttt{ANIMALS} category, and AudioSet's animal-related labels into a single unified corpus---a task that previously required manual tag reconciliation.

\subsection{CSV-Based Tag-to-UCS Conversion Pipeline}
\label{ssec:conversion}

The overall process for converting tag annotations to UCS \mbox{(sub)}\-categories is illustrated in Figure~\ref{fig:pipeline}.
Each input file carries one or more tags.
The pipeline classifies each tag independently through a four-stage cascade, collects the results, and then resolves conflicts at the file level when different tags map to different categories.

\begin{figure}
\centerline{\includegraphics[width=\columnwidth]{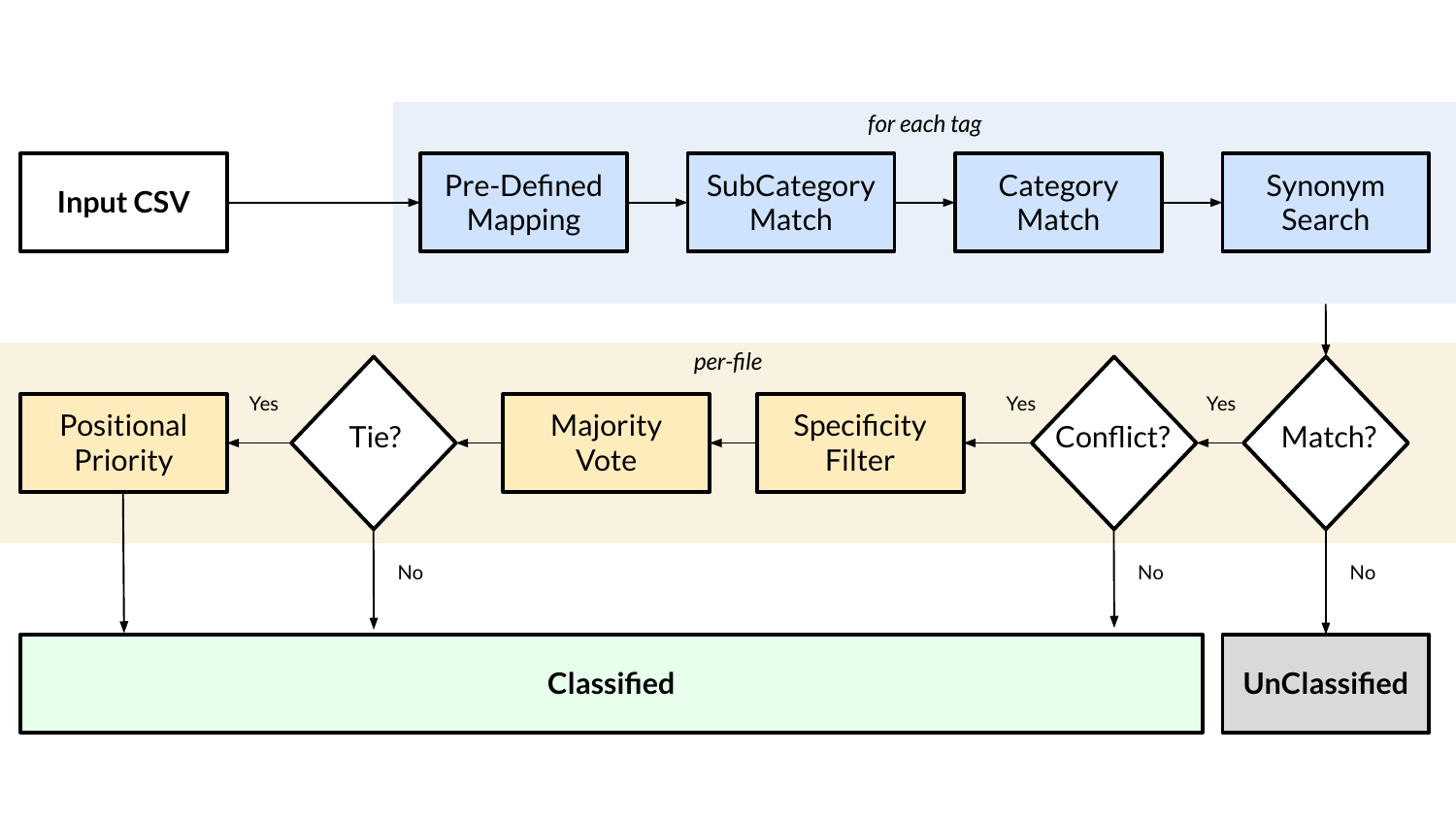}}
\caption{Tag-to-UCS conversion pipeline. \emph{Blue region (for each tag):} each tag passes through a four-stage cascade; the first matching stage classifies the tag. \emph{Yellow region (per file):} if at least one tag matched, the file proceeds to conflict resolution; otherwise it is marked unclassified.}
\label{fig:pipeline}
\end{figure}

\subsubsection{Per-tag classification}
\label{sssec:per_tag}

Each input tag is first normalized (lowercased, underscores replaced by spaces) and then passed through a four-stage cascade.
The stages are attempted in order; as soon as one stage produces a match, the tag is considered classified and the remaining stages are skipped:
\begin{enumerate}
	\item \textbf{Pre-defined Mapping}: a curated mapping table converts tags with dataset-specific conventions to UCS labels (e.g., {\texttt{gunshot\_and\_gunfire} $\rightarrow$ \texttt{GUNS/GUNSHOT}}).
	\item \textbf{SubCategory match}: the tag is matched against UCS subcategory names; if matched, the parent category is automatically derived from the UCS hierarchy.
	\item \textbf{Category match}: the tag is matched directly against UCS category names, yielding a category-level assignment without a subcategory.
	\item \textbf{Synonym match}: the tag is matched against the UCS synonym table ($9{,}972$~entries) via reverse lookup; if matched, both the category and subcategory are derived.
\end{enumerate}
If none of the four stages produces a match, the tag remains unclassified.

A file is classified as long as \emph{at least one} of its tags matches; it is marked unclassified only when \emph{none} of its tags match at any stage.
Tags that failed to match are retained in the metadata but do not influence the category assignment.

\subsubsection{Per-file conflict resolution}
\label{sssec:conflict}

If all matched tags for a file agree on the same UCS category, that category is assigned directly.
If matched tags disagree---i.e., different tags mapped to different categories---three rules are applied in sequence to select the final assignment:
\begin{enumerate}
	\item \textbf{Specificity filter}: tags that matched at the subcategory level are prioritized over category-only matches, as a subcategory match carries more precise semantic information. For example, if one tag maps to \texttt{GUNS/GUNSHOT} (subcategory level) and another maps to \texttt{DESIGNED} (category level only), the subcategory-level match is preferred.
	\item \textbf{Majority vote}: among the remaining matched tags, the category supported by the largest number of tags is selected.
	\item \textbf{Positional priority}: if a tie remains, the category associated with the last (rightmost) tag in the original annotation list is chosen. In FSD50K, where label propagation places broader tags after specific descriptors, later tags tend to represent the primary acoustic content rather than incidental sounds. For example, a clip tagged \texttt{Coin\_dropping, Singing, Human\_voice} is assigned to VOICES rather than FOLEY. AudioSet uses a different tag ordering convention; the applicability of this heuristic to AudioSet is acknowledged as a limitation (Section~\ref{ssec:limitations}). All files where this rule was applied are added to the ambiguity review list.
\end{enumerate}
Every file with at least one matched tag receives a final category assignment through these rules; no file is left without a category.
Files where matched tags produced competing categories are additionally logged to an ambiguity review list, which serves as an optional reference for manual verification.

\subsubsection{Auxiliary outputs}
\label{sssec:refinement}

The pipeline produces three auxiliary outputs: a list of unclassified files (i.e., files where all tags failed to match at any cascade stage), a frequency-ranked summary of unclassified tags, and the ambiguity review list.
High-frequency unclassified tags might be added by users as entries to the pre-defined mapping table in the configuration file to improve the conversion rate in a re-run.
All pipeline code, configuration files, and converted datasets are publicly available.\footnote{\label{fn:repos}Tools: \texttt{https://github.com/JunWooBeck/ucs-sfx-tools}\\\phantom{$^1$}Datasets: \texttt{https://github.com/JunWooBeck/fsd50k-ucs}, \texttt{https://github.com/JunWooBeck/audioset-ucs}, \texttt{https://github.com/JunWooBeck/esc50-ucs}, \texttt{https://github.com/JunWooBeck/envsound-ucs}.}

\subsection{UCS-aware dataset splitting}
\label{ssec:splitting}

Existing, previously proposed dataset splits do not account for the distribution within UCS categories. Using such existing splits might result in considerably different class distributions and imbalances between train and test sets. In the worst case, a training or test set class might have no entries.
Therefore, defining new splits for the UCS-annotated data is necessary. We provide a re-splitting tool that uses a composite stratification key formed by concatenating the Category and SubCategory fields (\texttt{Category||SubCategory}).
This preserves the joint distribution across both hierarchical levels.
The tool employs a two-pass splitting strategy: the first pass separates train+validation from test using stratified sampling, and the second pass divides train+validation into train and validation sets.
Category-SubCategory combinations with fewer than five samples cannot support stratified splitting and are therefore assigned entirely to the training set to prevent class absence in test or validation sets.
The splits are validated by checking for test/validation-only classes and computing the Pearson correlation between training and test category distributions, requiring $r > 0.99$ to confirm that the split preserves the original class balance.
Crucially, the tool supports merging multiple CSV files before splitting, directly enabling the aggregation of multiple datasets.
Default ratios are 70/15/15 (train/validation/test) with a fixed random seed for reproducibility.
The splitting tool additionally supports category-level filtering and multi-source merging through a JSON configuration file, enabling both aggregation of multiple datasets and extraction of targeted subsets (e.g., specifying only \texttt{WEATHER}, \texttt{WIND}, and \texttt{RAIN} to produce a focused outdoor ambience corpus).
Each file retains a \texttt{source\allowbreak\_\allowbreak dataset} field identifying its origin and preserves all original tags in the metadata, enabling post-hoc analysis of per-source contributions.

\section{Conversion results}
\label{ssec:conv_results}

\begin{table}
  \caption{Tag-to-UCS conversion results. AudioSet refers to the balanced training and evaluation segments only.}
  \centering
  \begin{tabular*}{\linewidth}{l@{\extracolsep{\fill}}rrrr}
    \toprule
    Dataset & Total & Classified & Rate & Ambiguous \\
    \midrule
    FSD50K   & 51{,}197 & 51{,}197 & 100\%   & 8{,}086 (15.8\%) \\
    AudioSet & 33{,}268 & 32{,}767 & 98.49\% & 9{,}160 (28.0\%) \\
    ESC-50   & 2{,}000  & 2{,}000  & 100\%   & 0 \\
    \bottomrule
  \end{tabular*}
  \label{tab:conversion}
\end{table}

Table~\ref{tab:conversion} summarizes the tag-to-UCS mapping results for the three well-known datasets FSD50K, AudioSet, and ESC-50.
We process FSD50K (dev and eval splits), AudioSet (balanced training and evaluation segments), and ESC-50.
The pipeline maps 100\% of FSD50K and ESC-50 files and 98.49\% of AudioSet files to UCS categories.
Because each file may carry multiple tags, a single successful tag match is sufficient to map the entire file; this multi-tag design accounts for the high file-level mapping rates despite individual tags sometimes failing to match.
ESC-50, with its compact 50-class vocabulary, achieves complete mapping without ambiguity.

Ambiguous files---those where multiple tags mapped to competing categories and required conflict resolution---account for 15.8\% of FSD50K and 28.0\% of AudioSet.
All ambiguous files receive a final category assignment through the conflict resolution rules described in Section~\ref{sssec:conflict}; the ambiguity review list is provided for optional manual inspection.
AudioSet's higher ambiguity rate reflects the fact that nearly every clip carries broad secondary tags such as \emph{Speech} and \emph{Music} regardless of the primary acoustic content, producing formal category conflicts that are resolved by the three-rule conflict resolution.

The 501 unmapped AudioSet files (1.51\%) contain only tags that either describe recording context rather than acoustic content (e.g., \texttt{Environmental\_noise}, \texttt{Heavy\_engine}) or use compound labels not covered by the mapping table (e.g., \texttt{Chopping\_\allowbreak(food)}, \texttt{Change\_\allowbreak ringing\_\allowbreak(bellringing)}).
These tags could be considered for addition to the mapping table for future conversion runs.

At the file level, the mapping table (Stage~1) resolves the vast majority of files: 97.0\% for FSD50K and 91.0\% for AudioSet.
Adding subcategory matching (Stage~2) raises coverage to 98.5\% and 96.1\%, respectively.
Category matching (Stage~3) and synonym matching (Stage~4) capture the remaining files, bringing FSD50K to 100\% and AudioSet to 98.5\% file-level coverage.
ESC-50 achieves 100\% mapping, with 86\% of files resolved at Stage~1 and the remainder via synonym matching (Stage~4).
These results confirm that the curated mapping table is the single most impactful component of the pipeline, while the automated matching stages provide incremental but essential coverage for long-tail tags.

\begin{figure*}[t]
\center
\includegraphics[width=\textwidth]{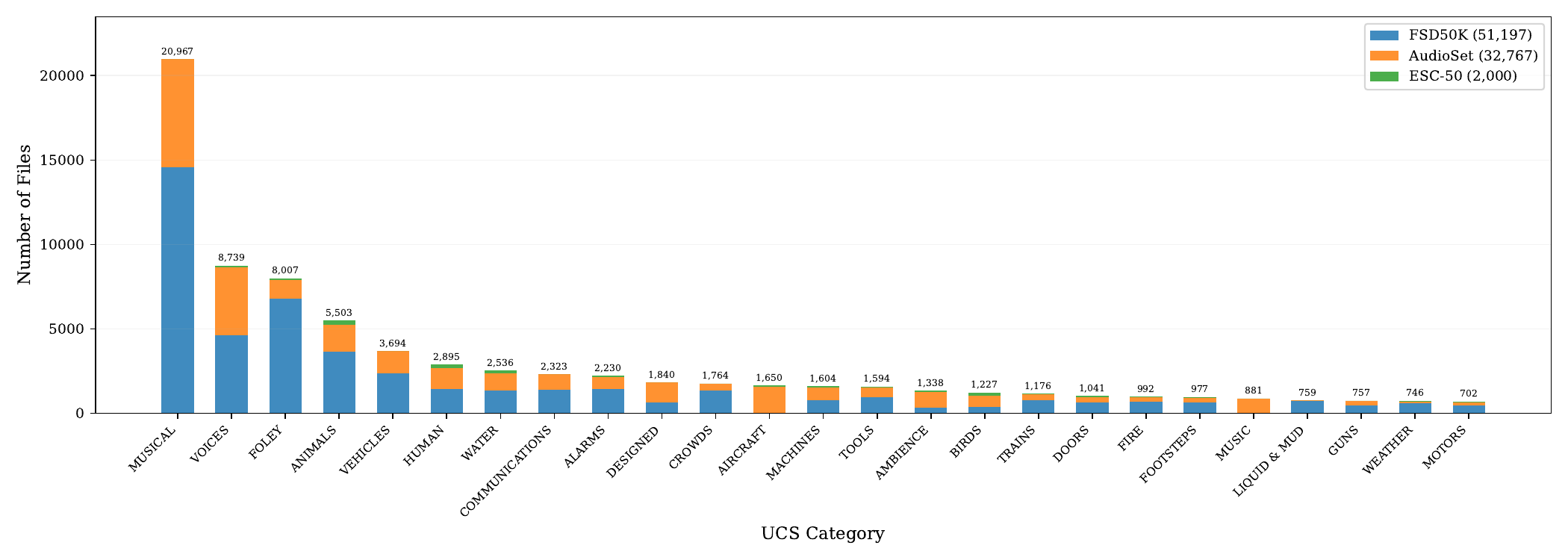}
\caption{UCS category distribution across FSD50K, AudioSet, and ESC-50 (top 25 categories by combined count). Each bar shows the per-source contribution.}
\label{fig:distribution}
\end{figure*}

Figure~\ref{fig:distribution} compares the UCS category distributions of the three datasets.
The long-tailed distribution is evident: \texttt{MUSICAL} alone accounts for over 20{,}000 files across FSD50K and AudioSet, while the smallest categories contain fewer than 700 files.
FSD50K and AudioSet show complementary coverage patterns, and ESC-50's 26 categories are a strict subset of both larger datasets.

\subsection{EnvSound-UCS: Environmental Sound Corpus}
\label{ssec:case_study}

To demonstrate the framework's practical capabilities and to provide consistent multi-source data for future research into SFX classification and generation, we present an environmental sound corpus by combining ESC-50, FSD50K, and AudioSet: \textit{EnvSound-UCS}.

\subsubsection{Configuration}
The same environmental sound filter is applied independently to each source dataset: the \texttt{MUSICAL} category is excluded (instrumental sounds not typically considered environmental), and the \texttt{VOICES} category is restricted to the \texttt{CRYING} and \texttt{LAUGH} subcategories only, as these correspond to non-verbal vocal events present in ESC-50 (\texttt{crying\_baby}, \texttt{laughing}).
The \texttt{MUSIC} category (referring to recorded background music, distinct from \texttt{MUSICAL} which covers instrument performance) is retained.
ESC-50, being an environmental sound dataset, contains no \texttt{MUSICAL} files and is therefore unaffected by the filter.

After filtering, each source is independently split into 70/15/15 train/validation/test partitions using the stratified splitting tool (Section~\ref{ssec:splitting}).
The per-source splits are then concatenated to form the final EnvSound-UCS corpus.
This split-then-merge strategy ensures that each source's internal distribution is preserved and, crucially, that the individual filtered datasets (FSD50K-env, AudioSet-env) can serve as standalone benchmarks for direct comparison with the combined corpus.

\subsubsection{Corpus statistics}
From a combined pool of 85{,}964 files, filtering produces 58{,}057 samples spanning 59 UCS categories: 33{,}065 from FSD50K (after \texttt{MUSICAL} removal), 22{,}992 from AudioSet (after \texttt{MUSICAL} removal), and 2{,}000 from ESC-50.

\begin{table}[t]
\caption{Source contributions for EnvSound-UCS (top 10 categories by sample count).}
  \centering
  \small
  \setlength{\tabcolsep}{5pt}
  \begin{tabular*}{\columnwidth}{lrrrr}
    \toprule
    Category & ESC & FSD & AS & Total \\
    \midrule
    \texttt{FOLEY}       & 120   & 6{,}814 & 1{,}073 & 8{,}007 \\
    \texttt{ANIMALS}     & 240   & 3{,}641 & 1{,}622 & 5{,}503 \\
    \texttt{VEHICLES}    & --    & 2{,}367 & 1{,}327 & 3{,}694 \\
    \texttt{HUMAN}       & 200   & 1{,}444 & 1{,}251 & 2{,}895 \\
    \texttt{WATER}       & 160   & 1{,}336 & 1{,}040 & 2{,}536 \\
    \texttt{COMMUNICATIONS} & -- & 1{,}419 & 904     & 2{,}323 \\
    \texttt{ALARMS}      & 80    & 1{,}422 & 728     & 2{,}230 \\
    \texttt{DESIGNED}    & --    & 669     & 1{,}171 & 1{,}840 \\
    \texttt{VOICES}      & 80    & 1{,}065 & 654     & 1{,}799 \\
    \texttt{CROWDS}      & --    & 1{,}349 & 415     & 1{,}764 \\
    \midrule
    \multicolumn{1}{l}{\emph{All 59 cats}} & 2{,}000 & 33{,}065 & 22{,}992 & 58{,}057 \\
    \bottomrule
  \end{tabular*}
  \label{tab:source}
\end{table}

Table~\ref{tab:source} breaks down the per-category contributions from each source.
FSD50K and AudioSet exhibit complementary coverage: for example, \texttt{ANIMALS} has 3{,}641~files in FSD50K but only 1{,}622 in AudioSet, while \texttt{DESIGNED} contains 669 FSD50K files but 1{,}171 from AudioSet.
Of the 59 categories, 33 have no ESC-50 representation (e.g., \texttt{COMMUNICATIONS}, \texttt{DESIGNED}, \texttt{CROWDS}), meaning that aggregation with larger datasets is the only way to obtain coverage in these domains.

\section{Benchmark experiment}
\label{sec:benchmark}

To establish the new dataset(s), we present benchmark results for the new categories and the new splits, as well as for the new combined environmental sounds dataset. We provide a detailed analysis of the results within the hierarchical label structure.

\subsection{Experiments}
\label{ssec:exp_setup}

We structure the benchmark into three experiments. The number of categories~(C) and subcategories~(S) varies across datasets (reported in the table headers). All models are trained and evaluated on 70/15/15 stratified splits. The framework's unified label space and provenance tracking enable controlled cross-source comparisons.

\textbf{Exp.~1: Flat classification.}
We train a direct category classifier ($Cat$) and a flat subcategory classifier ($SubCat_\mathit{flat}$). From the subcategory predictions, we derive category-level results ($Cat_\mathit{flat}$) by mapping predicted subcategories to their parent categories.
Each dataset uses its own train/test split.

\textbf{Exp.~2: Hierarchical classification.}
We train a hierarchical cascade classifier ($SubCat_\mathit{hier}$), where a category classifier routes each sample to a dedicated per-category subcategory classifier.
We additionally evaluate an oracle variant ($SubCat_\mathit{hier,orac}$) that uses ground-truth category labels for routing, establishing the upper bound of the hierarchical approach.

\textbf{Exp.~3: Cross-dataset evaluation.}
To investigate whether aggregating multiple training sources under UCS improves generalization, we train a model on the combined EnvSound-UCS training set and evaluate it on each source's individual test set.

\subsection{Classifier models and parametrization}
Audio features are extracted using PANNs CNN14~\cite{kong_panns_2020}, a pre-trained convolutional neural network that yields 2048-dimensional embeddings for each audio clip.
All classifiers share an identical architecture: a single linear layer projecting the 2048-dimensional embedding directly to the output classes, preceded by dropout ($p = 0.3$).
This uniform design ensures that performance differences across experiments reflect the classification strategy rather than architectural variation.
The setup is intentionally simple to establish an easily reproducible baseline.

All classifiers are trained with AdamW (learning rate $10^{-3}$, weight decay $10^{-3}$), cosine annealing learning rate scheduling, early stopping with patience of 20 epochs, and a maximum of 200 epochs.
To address class imbalance, we use focal loss~($\gamma = 2.0$) with inverse-frequency class weighting.
Identical hyperparameters are used for all datasets.
Each experiment is repeated with five random seeds (42, 123, 456, 789, 1024) and we report mean $\pm$ standard deviation.
The primary evaluation metric is the macro-averaged F1 score, which weights all classes equally regardless of their sample count and thus reflects performance on rare categories.

Note that PANNs CNN14 is pretrained on the full AudioSet training corpus ($\sim$2M clips).
For the AudioSet-UCS and EnvSound-UCS benchmarks, a portion of the test files originate from AudioSet's balanced training subset and were therefore seen during PANNs pretraining.
While the UCS category labels used here differ from AudioSet's original 527-class labels, the learned representations may still benefit from prior exposure to these audio files.
No part of FSD50K and ESC-50 were part of the PANNs training data.

\subsection{Results and Discussion}
\label{ssec:results}

\begin{table}
  \caption{Macro F1, mean $\pm$ std (5 seeds) on original datasets for their dataset-specific 70/15/15 split. Header notation~(C/S) indicates the number of categories~C and subcategories~S.}
  \centering
  \begin{tabular*}{\columnwidth}{l@{\extracolsep{\fill}}ccc}
    \toprule
     & FSD50K & AudioSet & ESC-50 \\
     & (37/59) & (60/158) & (26/20) \\
    \midrule
    $Cat$                        & .52{\tiny$\pm$.00} & .42{\tiny$\pm$.00} & .89{\tiny$\pm$.00} \\
    $Cat_\mathit{flat}$          & .71{\tiny$\pm$.00} & .56{\tiny$\pm$.00} & .99{\tiny$\pm$.00} \\
    $SubCat_\mathit{flat}$       & .65{\tiny$\pm$.00} & .49{\tiny$\pm$.00} & .95{\tiny$\pm$.01} \\
    $SubCat_\mathit{hier}$       & .60{\tiny$\pm$.00} & .42{\tiny$\pm$.00} & .95{\tiny$\pm$.01} \\
    $SubCat_\mathit{hier,orac}$ & .86{\tiny$\pm$.00} & .74{\tiny$\pm$.01} & .95{\tiny$\pm$.01} \\
    \bottomrule
  \end{tabular*}
  \label{tab:classification}
\end{table}

\begin{table}
  \caption{Macro F1, mean $\pm$ std (5 seeds) on environmental sound dataset classification with \texttt{MUSICAL} excluded and \texttt{VOICES} restricted to \texttt{CRYING} and \texttt{LAUGH}.
  Header notation~(C/S) indicates the number of categories~C and subcategories~S. Upper section: self-trained models.
  Lower section: model trained on the combined EnvSound-UCS corpus, evaluated on each source's test set.}
  \centering
  \small
  \setlength{\tabcolsep}{3pt}
  \begin{tabular*}{\columnwidth}{@{\extracolsep{\fill}}lcccc}
    \toprule
     & FSD-env & AS-env & ESC-50 & EnvSound \\
     & (36/50) & (59/142) & (26/20) & (59/144) \\
    \midrule
    \multicolumn{5}{l}{\emph{Self-trained}} \\ [3mm]
    $Cat$                        & .53{\tiny$\pm$.00} & .47{\tiny$\pm$.00} & .89{\tiny$\pm$.00} & .46{\tiny$\pm$.00} \\
    $Cat_\mathit{flat}$          & .72{\tiny$\pm$.00} & .57{\tiny$\pm$.00} & .99{\tiny$\pm$.00} & .55{\tiny$\pm$.00} \\
    $SubCat_\mathit{flat}$       & .66{\tiny$\pm$.00} & .49{\tiny$\pm$.01} & .95{\tiny$\pm$.01} & .49{\tiny$\pm$.00} \\
    $SubCat_\mathit{hier}$       & .61{\tiny$\pm$.00} & .42{\tiny$\pm$.01} & .95{\tiny$\pm$.01} & .39{\tiny$\pm$.00} \\
    $SubCat_\mathit{hier,orac}$ & .88{\tiny$\pm$.00} & .74{\tiny$\pm$.02} & .95{\tiny$\pm$.01} & .73{\tiny$\pm$.01} \\
    \midrule
    \multicolumn{5}{l}{\emph{Trained on EnvSound-UCS}} \\ [3mm]
    $SubCat_\mathit{flat}$       & .62{\tiny$\pm$.00} & .46{\tiny$\pm$.00} & .92{\tiny$\pm$.00} & \\
    \bottomrule
  \end{tabular*}
  \label{tab:env}
\end{table}

\subsubsection{Exp.~1: Flat classification}

The flat classification results $Cat$ are presented in Tables~\ref{tab:classification} and~\ref{tab:env}.
ESC-50 (26~categories) can be classified with a F1 of~.89, compared to~.52 for FSD50K (37~categories) and~.42 for AudioSet (60~categories).
Performance scales inversely with the number of classes; this relationship is expected, as classifying 60~categories is inherently harder than 26, and must be considered when comparing results across datasets.
The flat subcategory classification $SubCat_\mathit{flat}$ yields, interestingly, higher macro F1 across all datasets. 
Removing the \texttt{MUSICAL} category (Table~\ref{tab:env}, upper section) has only a negligible effect on performance, likely because \texttt{MUSICAL} is well-separated and acoustically distinct from the remaining categories.
Across all datasets, $Cat_\mathit{flat}$---the category-level F1 derived from subcategory predictions---consistently exceeds the direct category classifier $Cat$ (e.g., .71 vs.\ .52 on FSD50K, .56 vs.\ .42 on AudioSet). This suggests that the top-level UCS categories are broad and internally heterogeneous, making them difficult to model directly with a linear classifier. Learning at the finer subcategory granularity, where classes can be assumed to be more acoustically homogeneous, and then aggregating to categories yields better category-level discrimination.

\subsubsection{Exp.~2: Hierarchical classification}

On all datasets, the flat subcategory classifier outperforms the cascaded approach: $SubCat_\mathit{hier}$ trails $SubCat_\mathit{flat}$ by~5 percentage points on FSD50K, 7 on AudioSet, and~10 on EnvSound-UCS. ESC-50 shows virtually no difference between the two approaches.
This comparably poor performance is explained by the insufficient accuracy of the current category classifiers, which achieve F1 scores of only .42--.52 on the larger datasets. At these accuracy levels, routing errors propagate and impact the subcategory stage, outweighing the search space reduction benefits of hierarchical classification.

However, oracle routing reveals the potential of utilizing the hierarchy: $SubCat_\mathit{hier,orac}$ outperforms the flat baseline by~+20 percentage points on FSD50K, +25 on AudioSet-env, and +24 on EnvSound-UCS. The result for ESC-50 exhibits no meaningful oracle improvement, likely because the flat classifier already achieves a high F1 of~0.95 and the limited number of subcategories~(20) does not benefit from hierarchical routing.
This significant increase on the larger datasets indicates that improving category classifier accuracy could be key to unlocking the cascade's potential.

\begin{figure*}[!t]
\center
\includegraphics[width=\textwidth]{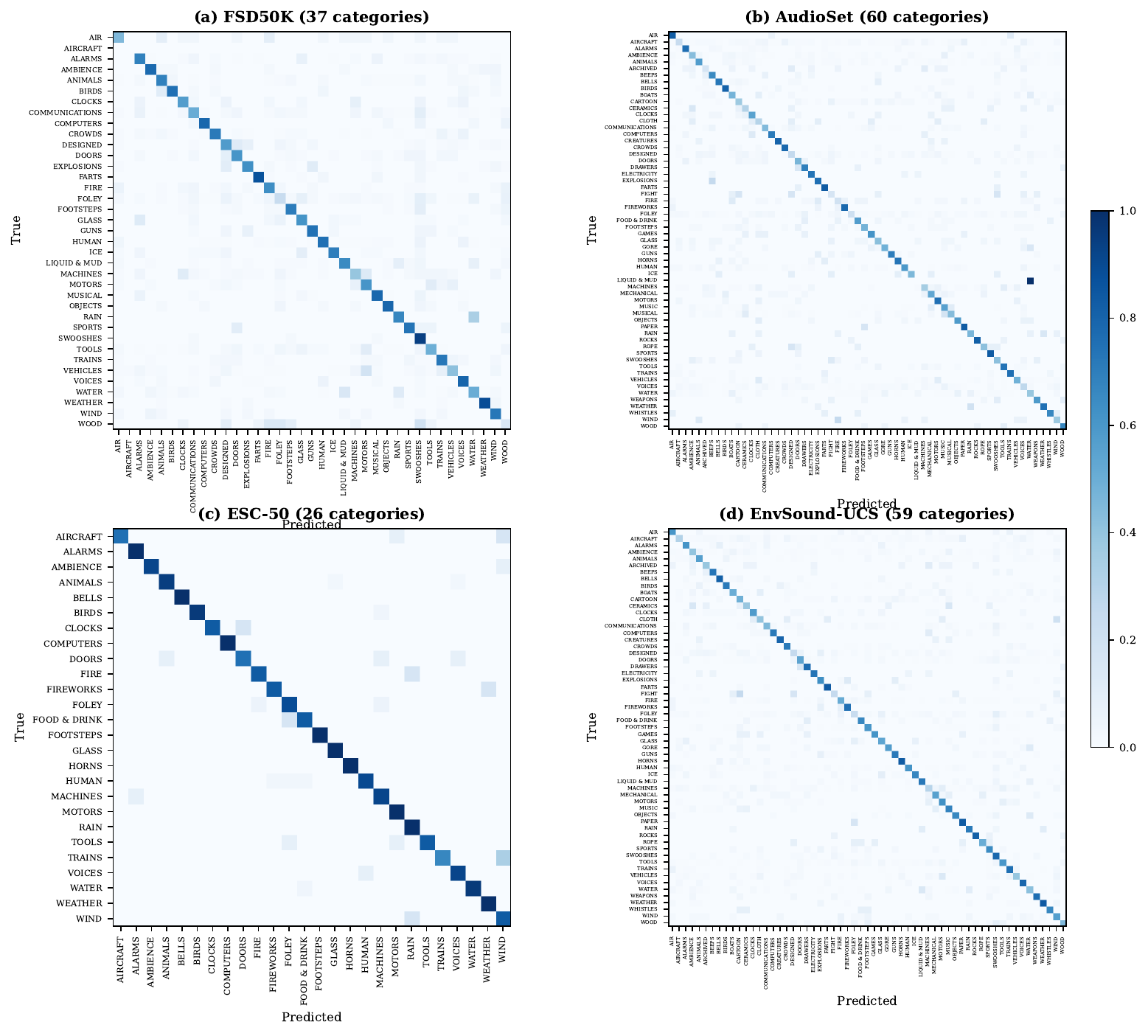}
\caption{Row-normalized category confusion matrices for (a) FSD50K, (b) AudioSet, (c) ESC-50, and (d) EnvSound-UCS. 
}
\label{fig:confusion}
\end{figure*}

Figure~\ref{fig:confusion} shows the category-level confusion matrices. Recurring off-diagonal confusion patterns include \texttt{VEHICLES}$\leftrightarrow$\texttt{MOTORS} and \texttt{LIQUID~\&~MUD}$\leftrightarrow$\texttt{WATER}.
These misclassifications follow acoustically motivated patterns: \texttt{VEHICLES} and \texttt{MOTORS} share engine-related sounds, making them inherently ambiguous at the category level, and \texttt{LIQUID~\&~MUD} and \texttt{WATER} both involve fluid dynamics, potentially having similar characteristics.

\subsubsection{Exp.~3: Cross-source evaluation}

The lower section of Table~\ref{tab:env} reports the cross-source results.
When the EnvSound-UCS model is evaluated on each source's test set, it trails the corresponding self-trained model by~4 percentage points on FSD50K-env, 3 on ESC-50, and~3 on AudioSet-env.

This performance degradation raises the question of why more training data fails to improve results.
A per-category analysis reveals that the primary cause is not class imbalance: the correlation between training distribution shift and per-category F1 change is not statistically significant for FSD50K-env (Pearson $r = -0.26$, $p = 0.13$), although a weak negative correlation is observed for AudioSet-env ($r = -0.35$, $p < 0.01$).

Instead, the following factors might explain the degradation.
First, the number of training classes increases when sources are combined: FSD50K-env has 36~categories, but the EnvSound-UCS model must classify into 59---including 23 categories absent from FSD50K.
This expanded output space increases training complexity without being directly relevant during inference.
AudioSet-env (59~categories) experiences minimal class expansion (59$\rightarrow$59), consistent with its smaller performance drop of only 3\%.

Second, dataset homogeneity might be a problem where the curated data especially of smaller datasets like ESC-50 may not always reflect the variety of input sources of real-world settings.
AudioSet, on the other hand, contains in-the-wild YouTube clips with variable recording conditions. When these heterogeneous sources are mixed, the model encounters less homogeneous acoustic representations for the same category.
The ESC-50 category \texttt{GLASS} illustrates this effect: the per-category F1 drops from .92 (self-trained) to .25 (EnvSound-trained), suggesting that the clean, curated ESC-50 class might be more challenging to model when including noisier representations from other datasets.

Finally, there is a chance that data and annotation quality might impact results. AudioSet, in particular, is known to include quality-impaired audio samples and potentially noisy annotations~\cite{fonseca_addressing_2020}. Adding noisy labels or audio data to the training set might thus negatively impact inference performance.

However, the evidence is inconclusive as to whether a more sophisticated classifier or re-training the feature extractor from scratch might benefit from the larger training dataset and yield better results with respect to generalizability and accuracy.

\section{Licensing Considerations}
\label{sec:licensing}

Combining multiple datasets under a unified taxonomy raises licensing questions worth considering.
We distinguish between \emph{audio} licenses (governing the sound files themselves) and \emph{metadata} licenses (governing annotations, ontologies, and derived labels).

\textbf{Audio licenses.}
The three source datasets carry different audio licenses.
FSD50K audio clips are individually licensed under various Creative Commons terms (CC0, CC-BY, CC-BY-NC, or CC~Sampling+), with per-clip information available in the dataset metadata.
ESC-50 audio is distributed under CC-BY-NC~3.0, prohibiting commercial use.
AudioSet does not distribute audio directly; it provides YouTube video identifiers and time segments, with the original audio subject to each video's individual copyright.

\textbf{Metadata licenses.}
FSD50K annotations are released under CC-BY~4.0.
AudioSet annotations are CC-BY~4.0, but the AudioSet ontology (class hierarchy and labels) is CC-BY-SA~4.0, meaning derivative works must be released under the same license.
Since our tag-to-UCS mapping table derives from the AudioSet ontology, it may be subject to the SA (ShareAlike) condition.
The UCS taxonomy itself is public domain, imposing no restrictions.
Our derived metadata (UCS labels, split files, and mapping tables) are released under CC-BY-SA~4.0.

\textbf{Implications for EnvSound-UCS.}
Because EnvSound-UCS aggregates sources with different licenses, users who assemble the full audio corpus must respect the most restrictive terms: ESC-50 audio carries a non-commercial (NC) constraint, and AudioSet audio cannot be redistributed due to YouTube Terms of Service.
Our framework distributes only CSV metadata files (UCS labels, splits, and conversion code), not audio files for any dataset, which mitigates redistribution concerns.
Users must independently obtain audio from the original sources: Freesound/Zenodo for FSD50K, YouTube for AudioSet, and GitHub for ESC-50.
Full per-dataset licensing details are provided in the datasheets (Appendix).

\section{Conclusion}
\label{sec:conclusion}

We have presented a UCS-based framework for standardized conversion and hierarchical annotation of sound effects datasets.
The CSV-based tag-to-UCS conversion pipeline achieves automatic mapping rates of 100\% on FSD50K and ESC-50, and 98.49\% on AudioSet.
The UCS-aware re-splitting tool preserves category and subcategory distributions through composite stratification keys and supports multi-source merging, as demonstrated by the construction of a 58{,}057-sample environmental sound corpus from three sources after category filtering (Section~\ref{ssec:case_study}).
Beyond aggregation, the framework's category-level filtering enables the extraction of targeted subsets (e.g., a focused outdoor ambience corpus from \texttt{WEATHER}, \texttt{WIND}, and \texttt{RAIN} categories), addressing the core goals outlined in the introduction: compatibility and interoperability through a shared label space, extensibility through a reusable conversion pipeline, and coherence through the hierarchical UCS structure that supports both broad and focused dataset construction.

The benchmark experiments reveal three key findings. First, category-level F1 derived from subcategory predictions consistently exceeds direct category classification, suggesting that the top-level UCS categories are broad and internally heterogeneous, making finer-grained subcategory learning a more effective strategy even for coarse-level tasks.
Second, oracle routing yields +20--25\% improvements over flat baselines on the larger datasets, demonstrating the latent potential of hierarchical classification under accurate category routing. Third, naive aggregation of heterogeneous data sources does not automatically improve over single-source training.

The deliverables of this work are distributed across four GitHub repositories (see footnote~\ref{fn:repos}).
The tool repositories contain the conversion pipeline, re-splitting tool, and configuration files; the data repositories contain metadata only (per-file UCS labels, train/vali\-dation/test partitions, and ambiguity review lists). Audio files are not redistributed.
All code and converted datasets are publicly available.
By bringing an industry-standard taxonomy into the academic domain, this work bridges the gap between professional sound design practice and audio machine learning research. With this release we hope to facilitate future research focusing on environmental sounds that links academic research with industry needs.

\subsection{Limitations and Future Work}
\label{ssec:limitations}

Several limitations should be noted.
First, the conversion pipeline operates entirely on text-based tag matching and does not perform auditory verification of the assigned categories.
The resulting UCS labels inherit any noisy labels in the original source annotations.
Second, the positional priority tie-breaking rule is invoked for approximately 10--20\% of classified files across datasets.
All files where tie-breaking was applied are recorded in the ambiguity review list, enabling users to identify and manually verify these cases.
Third, UCS version updates may require re-mapping; all results reported here are based on UCS v8.2.1.
Fourth, the benchmark experiments use PANNs CNN14 embeddings pretrained on the full AudioSet training corpus~\cite{kong_panns_2020} ($\sim$2M clips).
For AudioSet-UCS, approximately 51\% of test files originate from AudioSet's balanced training subset and were therefore seen during PANNs pretraining; for EnvSound-UCS this figure is approximately 20\%. While the UCS category labels used here differ from AudioSet's original 527-class labels, the learned representations may still benefit from prior exposure to these audio files. FSD50K and ESC-50 are not part of the PANNs training data and are unaffected.
Finally, the current pipeline requires a manually curated mapping table to handle dataset-specific tag conventions.
While this table is reusable across subsequent conversions of the same dataset, extending the framework to new datasets requires adding new mappings.
Future work could leverage large language models to automate this mapping process, and domain-aware training strategies (e.g., source-weighted sampling or domain adaptation) could address the performance degradation observed in naive multi-source aggregation.
The potential of hierarchical classification demonstrated by the oracle experiments could be further explored through taxonomy-aware learning~\cite{liang_taxonomy_2024} and multimodal approaches~\cite{anastasopoulou_hierarchical_2025}.

\FloatBarrier
\section{Acknowledgments}
The authors thank the anonymous reviewers for their feedback and the UCS community for maintaining the taxonomy as a public resource.

\bibliographystyle{IEEEtranDAFx}
\bibliography{DAFx26_tmpl} 

\clearpage


\appendix
\section{Dataset Datasheets}
Following the framework proposed by Gebru et al.~\cite{gebru_datasheets_2021}, we provide datasheets for the four datasets used in this work.
Our repositories distribute only CSV metadata files; audio files must be obtained from the original sources.

\subsection{FSD50K (UCS-Mapped)}

\textbf{Motivation.}
\begin{compactitem}
\item \emph{Purpose:} Large-scale open dataset of human-labeled sound events~\cite{fonseca_fsd50k_2022}, re-labeled to UCS.
\item \emph{Creators:} Fonseca et al.\ (Music Technology Group, UPF). UCS re-labeling by the authors.
\end{compactitem}

\textbf{Composition.}
\begin{compactitem}
\item 51,197 classified files (40,966 dev + 10,231 eval), 100\% classification rate, 37 UCS categories
\item Variable-length audio clips (0.3--30s) from Freesound, re-mapped to UCS Category and SubCategory
\item Single UCS Category/SubCategory per file; original FSD50K tags retained
\item 70/15/15 stratified splits (seed~42)
\item 8,086 files flagged for ambiguity review; ties resolved deterministically without auditory verification
\end{compactitem}

\textbf{Collection.}
\begin{compactitem}
\item Audio from Freesound.org (Creative Commons licensed); labels via automated retrieval + human validation
\item UCS labels derived algorithmically via the conversion pipeline
\end{compactitem}

\textbf{Uses \& Limitations.}
\begin{compactitem}
\item Used as UCS-relabeled benchmark for flat and hierarchical classification
\item UCS labels are algorithmically derived and should not be treated as verified ground truth
\end{compactitem}

\textbf{Distribution \& License.}
\begin{compactitem}
\item UCS labels and splits via GitHub repositories (see footnote~\ref{fn:repos}); audio from Zenodo
\item Audio: Creative Commons (varies per clip). Metadata: CC-BY-SA~4.0
\item Maintained by the authors; updates follow UCS revisions
\end{compactitem}

\subsection{AudioSet (UCS-Mapped Subset)}

\textbf{Motivation.}
\begin{compactitem}
\item \emph{Purpose:} Large-scale human-labeled audio event dataset~\cite{gemmeke_audio_2017}, balanced subset re-labeled to UCS.
\item \emph{Creators:} Gemmeke et al.\ (Google Research). UCS re-labeling by the authors.
\end{compactitem}

\textbf{Composition.}
\begin{compactitem}
\item 32,767 classified files (17,176 balanced\_train + 15,591 eval) + 501 unclassified (1.51\%), 60 UCS categories
\item 10-second clips from YouTube; multi-label AudioSet tags re-mapped to single UCS Category/SubCategory
\item Unbalanced training set ($\sim$2M) excluded due to computational constraints
\item 70/15/15 stratified splits (seed~42)
\item 9,160 files flagged for ambiguity review; original labels contain crowd-sourced noise
\end{compactitem}

\textbf{Collection.}
\begin{compactitem}
\item Audio from YouTube (Google Research); human-annotated with AudioSet ontology
\item UCS conversion pipeline: pre-defined mapping $\rightarrow$ SubCategory match $\rightarrow$ Category match $\rightarrow$ synonym lookup $\rightarrow$ conflict resolution
\end{compactitem}

\textbf{Uses \& Limitations.}
\begin{compactitem}
\item Used as UCS-relabeled benchmark for flat and hierarchical classification
\item Not suitable as ground truth for UCS taxonomy evaluation; original AudioSet labels contain annotation noise
\end{compactitem}

\textbf{Distribution \& License.}
\begin{compactitem}
\item UCS labels and splits via GitHub repositories (see footnote~\ref{fn:repos}); audio from AudioSet website (YouTube ToS applies)
\item AudioSet annotations: CC-BY~4.0; ontology: CC-BY-SA~4.0. Our metadata: CC-BY-SA~4.0
\item Maintained by the authors
\end{compactitem}

\subsection{ESC-50}

\textbf{Motivation.}
\begin{compactitem}
\item \emph{Purpose:} Environmental sound classification benchmark~\cite{piczak_esc_2015}, re-labeled to UCS.
\item \emph{Creators:} Karol J.\ Piczak. UCS re-labeling by the authors.
\end{compactitem}

\textbf{Composition.}
\begin{compactitem}
\item 2,000 clips, 50 classes, 5 major categories, 40 clips/class, 26 UCS categories
\item 5-second WAV files at 44.1\,kHz from Freesound.org; single categorical label + derived UCS Category
\item Original 5-fold CV available; 70/15/15 stratified split applied for EnvSound study
\item Curated dataset with clean labels; UCS mapping straightforward
\end{compactitem}

\textbf{Collection.}
\begin{compactitem}
\item Manually selected from Freesound.org; all clips Creative Commons licensed
\end{compactitem}

\textbf{Uses \& Limitations.}
\begin{compactitem}
\item Standalone benchmark + source dataset for EnvSound-UCS
\item Limited scope (50 classes, 2,000 clips); not suitable for comprehensive general-purpose evaluation
\end{compactitem}

\textbf{Distribution \& License.}
\begin{compactitem}
\item ESC-50 via GitHub under CC-BY-NC~3.0; UCS mappings via GitHub repositories (see footnote~\ref{fn:repos})
\item Maintained by Piczak (original) and the authors (UCS mappings)
\end{compactitem}

\subsection{EnvSound\_UCS}

\textbf{Motivation.}
\begin{compactitem}
\item \emph{Purpose:} Combined environmental sound benchmark from ESC-50 + FSD50K + AudioSet under UCS.
\item \emph{Creators:} The authors.
\end{compactitem}

\textbf{Composition.}
\begin{compactitem}
\item 58,057 samples across 59 UCS categories (\texttt{MUSICAL} excluded, \texttt{VOICES} restricted to \texttt{CRYING}/\texttt{LAUGH})
\item Union of all UCS-classified files from the three sources that map to retained categories
\item Per instance: UCS Category, SubCategory, original keywords, source provenance (CSV metadata only)
\item Split-then-merge: each source split 70/15/15 independently (seed~42), then concatenated
\item Inherits noise from all sources; cross-source label inconsistencies possible by design
\end{compactitem}

\textbf{Collection.}
\begin{compactitem}
\item No new audio collected; assembled via \texttt{combine\_and\_split.py} (load CSVs $\rightarrow$ filter categories $\rightarrow$ verify files $\rightarrow$ stratified split)
\end{compactitem}

\textbf{Uses \& Limitations.}
\begin{compactitem}
\item Combined benchmark: $SubCat_\mathit{flat}$ F1 = 0.49, $SubCat_\mathit{hier,orac}$ F1 = 0.73
\item Not suitable for evaluating UCS taxonomy quality; account for source-level differences in comparisons
\end{compactitem}

\textbf{Distribution \& License.}
\begin{compactitem}
\item Configuration and split CSVs via GitHub repositories (see footnote~\ref{fn:repos}); audio from original sources
\item Audio licenses: AudioSet (YouTube ToS), FSD50K (CC varies), ESC-50 (CC-BY-NC~3.0). Metadata: CC-BY-SA~4.0
\item Maintained by the authors; version-controlled for future source additions and UCS revisions
\end{compactitem}

\end{document}